\documentstyle[aps,floats,graphicx]{revtex}

\begin{document}

\newcommand{\bec}{\begin{center}}
\newcommand{\ec}{\end{center}}
\newcommand{\be}{\begin{equation}}
\newcommand{\ee}{\end{equation}}
\newcommand{\beqn}{\begin{eqnarray}}
\newcommand{\eeqn}{\end{eqnarray}}
\newcommand{\bet}{\begin{table}}
\newcommand{\ent}{\end{table}}
\newcommand{\bib}{\bibitem}

\wideabs{

\title{
Intermixing in Cu/Co: molecular dynamics simulations and 
Auger electron spectroscopy depth profiling
}

\author{M. Menyh\'ard, P. S\"ule}
  \address{Research Institute for Technical Physics and Materials Science,\\
Konkoly Thege u. 29-33, Budapest, Hungary,sule@mfa.kfki.hu,www.mfa.kfki.hu/$\sim$sule\\
}

\date{\today}

\maketitle

\begin{abstract}
The ion-bombardment induced evolution of intermixing is studied by molecular dynamics simulations
and by Auger electron spectroscopy depth profiling analysis (AESD) in Cu/Co multilayer.
It has been shown that
from AESD we can derive
the low-energy mixing rate and which can be compared with
the simulated values obtained by molecular dynamics (MD) simulations.
The overall agreement is reasonably good hence MD can hopefuly be used to estimate
the rate of intermixing in various interface systems.


\end{abstract}
}

  Low-energy ion beams are commonly used in surface analysis and for film growth \cite{Donnelly,IAFG}.
  The use of ion-sputtering in the controllable production of nanostructures and self-assembled nanoppaterns  have also become one of the most important fields
in materials science \cite{IAFG,Persson,Timm,Facsko}.

 As the dimensions of the surface features is reduced to the nanoscale, many classical macroscopic (continuum and mesoscopic) models for morphological evolution lose their validity.
Therefore the understanding of the driving forces and laws governing mass transport involved in the synthesis and organisation of nanoscale features in solid state materials
is inevitable.
Unfortunately the fundamental understanding and the nanoscale control of atomic transport processes, such as interdiffusion 
is not available yet \cite{Adamowich,Ladwig}.
In order to get more insights in the atomic relocation processes during postgrowth
low-energy ion-sputtering, it is important to measure and to calculate accurately the rate of
intermixing at the interface.

 In this Letter we will show that using a newly developed code for simulating ion-sputtering based on molecular dynamics
we are able to get mixing rates.
We will also show that it is possible to extract the mixing rate data from AESD.
It turns out that the agreement of the two methods is reasonably good.
In this way we also could check the reliability of molecular dynamics simulations.

 The AESD \cite{Hofmann,Barna} measurements were carried out on a Cu/Co multilayer system. The sample was made by sputter deposition on polished single-crystal (111) silicon substrates in a plasma beam sputter deposition system. It was characterized by XTEM and RBS and flat interfaces have been found \cite{Barna}. AESD depth profiling was carried out using a dedicated device \cite{Barna2} by applying Ar$^+$ ions of energy of 1 keV and angle of incidence of $10^{\circ}$ with respect to the surface of the crystal (grazing angle of incidence). The sample was rotated during ion bombardment. 

 The depth profiles were measured as a function of the sputtering time keeping the bombarding ion current constant. A STAIB DESA 100 pre-retarded CMA with fixed energy resolution was used to record the AES spectra. The following AES peaks were detected Cu (60 eV), Cu (920 eV) and Co (656 eV, to avoid overlapping). A part of a typical depth profile is shown in Fig 1. For clarity only the copper is shown in Fig 1. To calculate the concentration the intensity of the copper Auger current was normalized to that of the pure copper. The depth scale was calculated from the known thickness of the sample \cite{Barna}.

It is generally supposed that during ion mixing the atomic movements are similar to that of the usual diffusion and thus the occurring broadening can be described by the same equations \cite{Nastasi}. Accordingly in case of a bilayer system the concentration distribution formed due to ion bombardment of the interface by a given fluence can be described by the $erf$ function. The variance, $\sigma^2$, of the $erf$ function determines the extent of ion mixing.
 In many cases $\sigma^2$ depends linearily  on the ion fluence $\Phi$. In these cases the $\sigma^2/\Phi$ ratio called as mixing rate characterizes the mixing for a given ion bombarding condition. Important
advantage of using the term mixing rate is that it can be in principle directly derived from the experiments. 

  In the case of AESD we measure the Auger current of the elements present. The measured
Auger current of element $j$
 $I_j$ depends on the in-depth distribution as follows:
\begin{figure}[hbtp]
\begin{center}
\includegraphics*[height=5cm,width=6cm,angle=0.]{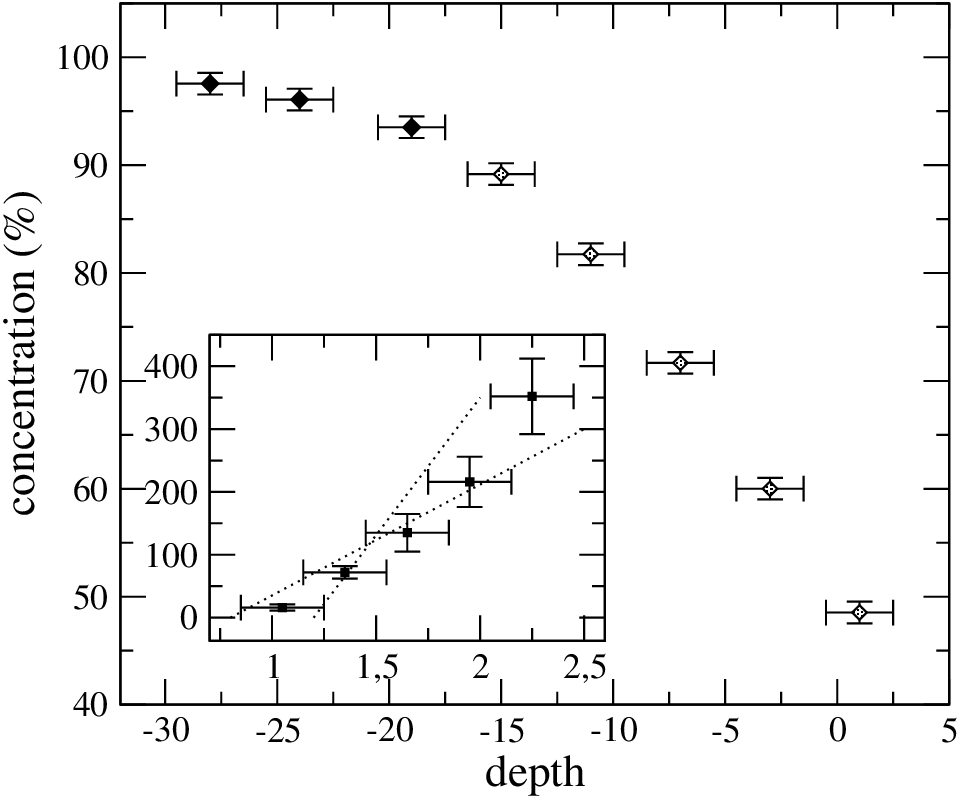}
\caption[]{
The measured depth profile obtained at 1 keV ion energy.
The concentration of Cu is given as a function of the removed
layer thickness (measured from the interface in $\hbox{\AA}$).
Empty diamonds denote the points at which $\sigma^2$ values are determined
by AESD.
{\em Inset figure}:
The broadening ($\sigma^2$) of the interface as a function of the fluence (ion/
$\hbox{\AA}^2$).
The dotted lines show the two extreme of the possible slopes (mixing rates).
}
\label{fig1}
\end{center}
\end{figure}
$I_j=i_j(1)+i_j(2) \kappa_j+i_j(3) \kappa_j^2+...$, where $i_j(k)$ is the Auger current of element $j$ emitted by the atomic layer $k$, and $\kappa_j$ gives the attenuation of the Auger current crossing an atomic plane. $i_j(k) \approx X_j(k)N(k)$, where $X_j(k)$ is the concentration of element $j$ in layer $k$, while $N(k)$ is the number of atoms  of the $k$-th atomic plane. It is evident that in general from a single measured $I_j$ one cannot determine the $i_j(k)$ values. On the other hand this equation can be used to simulate the measured Auger current during the depth profiling procedure if we assume an in-depth distribution.

 We do not know any experimental measurement of the in-depth distributions formed during AESD applying low ($0.2-2$ keV) ion energy. It seems, however, that dynamic TRIM simulation can reliably be used to describe AESD \cite{Trim}. This calculation also provides the in-depth distribution during the procedure. It turns out that at the beginning of the depth profiling procedure, when the interface is still far from the surface the in-depth distribution can be approximated by the $erf$ function. For the evaluation of the
experimentally determined depth profile we will suppose that
(i) the interface has an intrinsic surface roughness, which can also be approximated by an $erf$ function of $\sigma_0$. 
(ii) ion mixing is the only process contributing to the broadening of the interface at least in the beginning part of the depth profiling, which will be studied. 
Thus for any measured copper Auger current we should find $x_0$ (the distance of the interface from the surface) and $\sigma_m$ (the measured variance) values of the $erf$ function. Then the variance due to the ion bombardment induced mixing is $\sigma^2= \sigma_m^2-\sigma_0^2$.
We derived $\sigma^2$ at depths as indicated in Fig 1. at 1 keV ion energy at $\sim 10^{\circ}$ angle of incidence with respect to the surface. To proceed we must know the number of ions causing the broadening at the interface. In our experimental arrangement we cannot measure the ion fluence. On the other hand we can measure accurately the removed layer thickness. Taking the sputtering yield from the literature $Y \approx 1.2$ \cite{Barna}, we can derive the curve $\sigma^2$ vs. fluence $\Phi$ which is shown in inset Fig. 1. Inset Fig. 1 also 
\begin{figure}[hbtp]
\begin{center}
\includegraphics*[height=4.5cm,width=6cm,angle=0.]{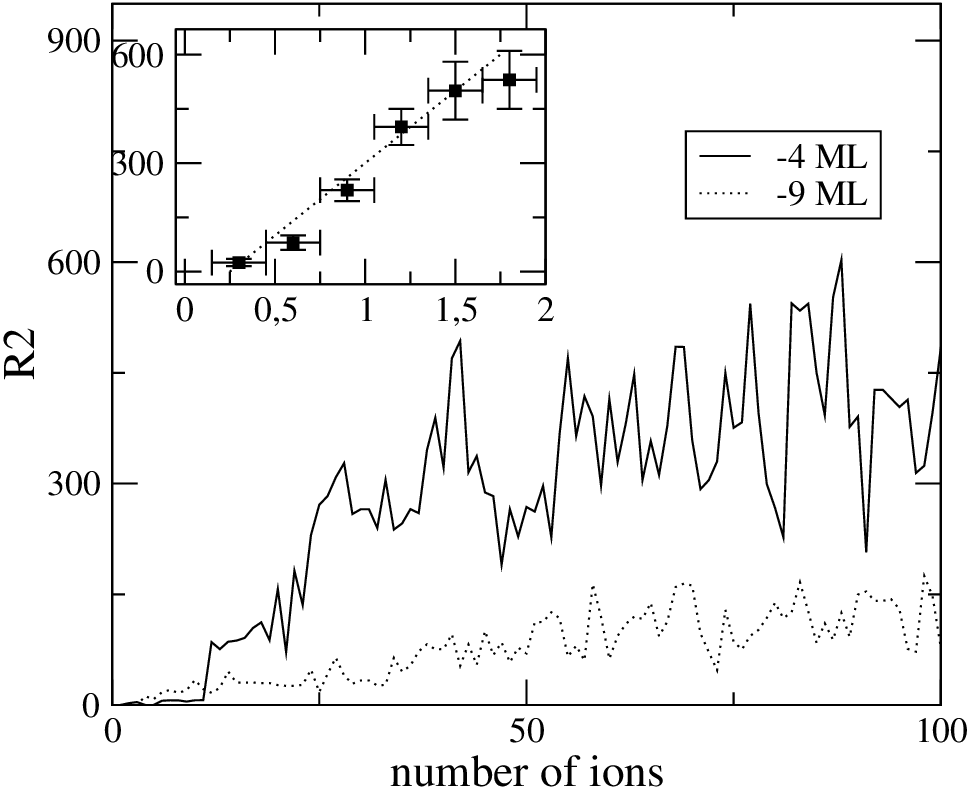}
\caption[]{
The simulated filtered $\langle R^2 \rangle$ (R2) as a function of the number of the ion impacts (number fluence)
at 1 keV ion energy at grazing angle of incidence using simulated ion-sputtering (molecular
dynamics).
Dotted line is a linear fit to the $\langle R^2 \rangle$ curves.
$\langle R^2 \rangle$ is simulated with ion impact depth positions at the free surface (9 ML above the interface) and also at $4$ ML above the interface.
{\em Inset figure}:
The simulated filtered $\langle R^2 \rangle$ for sputter removal
up to the removal of 4 layers as a function of the ion dose (ion/$\hbox{\AA}^2$).
The error bars are standard deviations.
}
\label{fig2}
\end{center}
\end{figure}
shows the two limiting slopes; thus we can derive the mixing rate being
$250 \pm 150$ $\hbox{\AA}^4$.

  Classical constant volume molecular dynamics simulations were also used to simulate the ion-solid interaction
(ion-sputtering) using the PARCAS code \cite{Nordlund_ref}.
Further details  are given in ref. 
\cite{Nordlund_ref,Sule_PRB05,Sule_NIMB04}.
We irradiate the bilayer Cu/Co (9 monolayers, (ML) film/substrate)
with 1 keV Ar$^+$ ions repeatedly with a time interval of 5-20 ps between each of
the ion-impacts at 300 K
which we find
sufficiently long time for the termination of interdiffusion, such
as sputtering induced intermixing (ion-beam mixing) \cite{Sule_NIMB04}.
 The initial velocity direction of the
impinging ions were $10^{\circ}$ with respect to the surface of the crystal (grazing angle of incidence).

 To describe homo- and heteronuclear interaction of Cu and Co, the Levanov's tight-binding
potentials are used \cite{Levanov}.
The cutoff radius $r_c$ is taken as the second neighbor distance.

 We randomly varied the impact position and the azimuth angle $\phi$.
In order to approach the real sputtering limit a large number of ion irradiation are
employed using simulations conducted subsequently together with analyzing
the history files (movie files) in each irradiation steps.
In this article we present results up to 100 ion irradiation which we find suitable for
comparing with low to medium fluence experiments. 100 ions are randomly distributed
over a $40 \times 40$ \hbox{\AA}$^2$ area.
The size of the simulation cell is $100 \times 100 \times 75$ $\hbox{\AA}^3$ including
62000 atoms.
The computer animations can be seen in our web page \cite{web}.

 The simulation of the sputter-removal has been carried out in the
following way:
 First we ion-bombarded the first (top) layer (the distance from the interface is -9 ML) 
using 100 ions, than using the ion-sputtered simulation cell again ion-sputtering is
simulated by 100 ions in the 2nd layer (-8 ML). Following this procedure in the 3rd,
4th and 5th layers we get an ion-sputtered sample with an intermixed interface.
We find that $\langle R^2 \rangle$ remains rather small until the ions are initilaized
in the 3-4th layers (-5 ML, see Fig. 2). This is because 
the distance from the interface is beyond the projected range of the 1 keV ions
in the upper layers.
To reduce the computational demand,
we do not remove completely the layers.
\begin{figure}[hbtp]
\begin{center}
\includegraphics*[height=4.5cm,width=5.5cm,angle=0.]{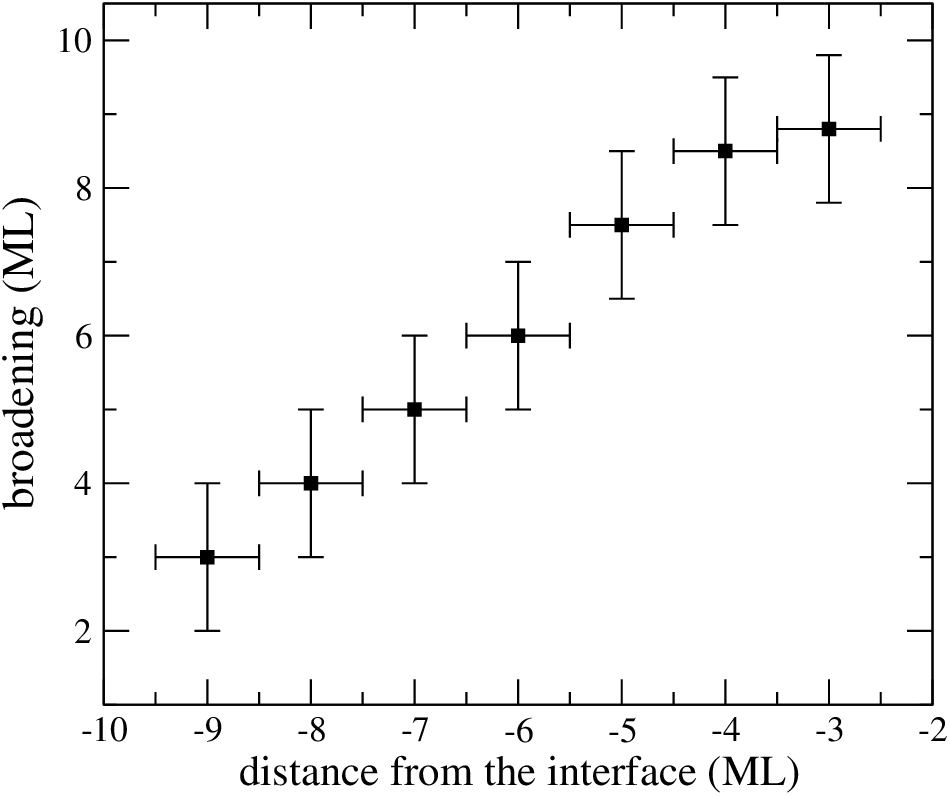}
\caption[]{
The simulated broadening (in ML) as a function of the distance from the interface (in ML)
at 1 keV ion energy at $10^{\circ}$ angle of incidence
obtained by simulated sputter-removal.
$-9$ ML corresponds to the free surface.
The error bars are standard deviations.
}
\label{fig3}
\end{center}
\end{figure}

  In Fig 2 the evolution of the sum of the square atomic displacements (R2) along the depth direction of all intermixing atoms $\langle R^2 \rangle= \sum_i^N [z_i(t)-z_i(t=0)]^2$, where $z_i(t)$ is the depth position of atom $i$ at time $t$, can be followed as a function of the 
number of ions.
Using the relation
$\sigma^2 = \langle R^2 \rangle$ it is possible to calculate the mixing rate
$k=\langle R^2 \rangle/\Phi$ from the slope of the fitted straigh line
on the $\langle R^2 \rangle$ vs. $\Phi$ curve in Fig 2.
The contribution of intermixed atoms to $\langle R^2 \rangle$ is excluded (filtered $\langle R^2 \rangle$) in a given layer  
 if the in-layer concentration of them is less then $\sim 16$ \%.
The filtered simulated $\langle R^2 \rangle$ is used to compare with the 
measured $\sigma$ values.
The lower concentratioan limit of $\sim 16$ \% corresponds
to the usual $84-16$ rule used for extracting the 
depth resolution in AESD \cite{depthresol}.
Therefore, the comparison with AESD is meaningful when 
contributions to broadening (and to $\langle R^2 \rangle$) are filtered out
which gives less in-layer concentartion than $\sim 16$ \%.
The results of simulated sputter-removal is shown in the inset Fig 2.
We also show results in Fig 2 for simplified simulations, when ions are directly initialized
from between the 4th and 5th layers (embedded ion-sputtering) using also grazing angle of incidence.
The evolution of the filtered $\langle R^2 \rangle$ as a function of the number of ions
shows us that $\langle R^2 \rangle$ is very close to that obtained for simulated
sputter removal.
Therefore we conclude from these results that it is sufficient to simulate 
intermixing with embedded ion-sputtering.

 The calculated mixing rate on the basis of inset Fig 2, is $k \approx 350 \pm 100$ $\hbox{\AA}^4$.
This value is reasonably comparable with the value obtained by AESD measurements.
It must also be noted that Cai {\em et al.} obtained
$k \approx 400$ $\hbox{\AA}^4$ using 1 MeV Si$^+$ ions and X-ray scattering techniques
in Co/Cu multilayer \cite{Cai}.
Since the ion mixing occur at the low energy end of the cascade process, e.g.
the high energy collisional cascades split into low energy subcascades (see e.g. refs. in \cite{Sule_PRB05}),
an agreement is expected for.
We also give the measured and calculated mixing efficiencies of $\xi =k/F_D$ \cite{Nastasi,Sule_PRB05}, where
$F_D$ is the deposited ion-energy/depth ($F_D \approx 45$ eV/$\hbox{\AA}$ obtained by SRIM \cite{SRIM} simulation).
We get $\xi
 \approx 6 \pm 3$ $\hbox{\AA}^5$/eV by experiment and $\xi
 \approx 7 \pm 2$ $\hbox{\AA}^5$/eV by simulations.
On the basis of this value we can characterize low-energy ion-sputtering induced intermixing in Cu/Co
as a ballistic interdiffusion process.

 The saturation of broadening is reached at around $\sim 9 \pm 1$ ML interface width as it is shown in Fig 3 obtained by simulated sputter-removal.

  In this Letter we have presented
that the combination of atomistic simulations with
Auger electron spectroscopy depth profiling might be a new efficient method
to depth profiling analysis of multilayered materials.
Also, the reasonably good agreement between experiment and simulations
provides us the possibility of predicting interdiffusion properties for various multilayers
for which no experimental results are available.

{
\scriptsize
This work is supported by the OTKA grant F037710
from the Hungarian Academy of Sciences.
The work has been performed partly under the project
HPC-EUROPA (RII3-CT-2003-506079) with the support of
the European Community using the supercomputing
facility at CINECA in Bologna.
The help of the NKFP project of
3A/071/2004 is also acknowledged.
}

\end{document}